\documentclass[%
 reprint,
 amsmath,amssymb,
 aps,
]{revtex4-1}
\usepackage{esvect}
\usepackage{rotating}
\usepackage{graphicx}
\usepackage{dcolumn}
\usepackage{bm}

\begin{document}

\preprint{APS/123-QED}

\title{Precise energy eigenvalues of hydrogen-like ion moving in quantum plasmas}

\author{S. Dutta}
\affiliation{%
 Belgharia Texmaco Estate School, Belgharia, Kolkata 700 056, India
}%


\author{J. K. Saha}
\affiliation{
 Indian Association for the Cultivation of Sciene, Jadavpur, Kolkata 700 032, India
}%
\author{T. K. Mukherjee}
\homepage{drtapanmukherjee@gmail.com}
\affiliation{%
 Narula Institute of Technology, Agarpara, Kolkata 700 109, India
}%


\date{\today}

\begin{abstract}
The analytic form of the electrostatic potential felt by a slowly moving test charge in quantum plasma is being derived. It has been shown that the potential composed of two parts: Debye-Huckel screening term and near-field wake potential which depends on the velocity of the test charge and the number density of the plasma electrons. Rayleigh-Ritz variational calculation has been done to estimate precise energy  eigenvalues of hydrogen-like ion under such plasma environment. A detailed analysis shows that the energy levels are gradually moves to the continuum with increasing plasma electron density while level crossing phenomenon have been observed with the variation of ion velocity.  
\begin{description}
\item[PACS numbers]
52.25.Vy, 52.40.-w, 31.15.xt
\end{description}
\end{abstract}

\pacs{Valid PACS appear here}
\maketitle


\section{\label{sec:level1}Introduction}
The study on the change in structural properties of foreign atoms or ions in different external environment [1-14] is a subject matter of immense interest in the last few decades as it provides deep insight to several interesting phenomenon in astrophysics and plasma physics. There exists a bulk of studies detailed in Sil \textit{et. al.} [15], on the behavioral changes in the structural properties of few-body systems embedded in external plasma environment, useful for laboratory and astrophysical plasma diagnostics determination. The most important part of such studies is to model the environment by an effective potential so that one can suppose as if the foreign atom/ion or the test charge will feel that effective potential while placed in or moving through that medium. It is well-known according to the Debye-Huckel theory [16, 17] of weak electrolyte that an static atom/ion feel screened Coulomb type potential while placed in a collision-less high temperature classical plasma. The screening parameter in this case is a function of electron number density ($n_{e}$) and temperature ($T$) of the plasma and thus different plasma situations can be simulated by suitably tuning the screening parameter [4]. In contrast, when the temperature ($T$) of the plasma electrons approaches to the ``\textit{Fermi-temperature}'' $T_{F}=E_{F}/k_{B}$ [$E_{F}$ being the ``\textit{Fermi-energy}'' of the electrons and $k_{B}$ is the Boltzmann constant], the equilibrium plasma electron distribution function changes from Maxwell-Boltzmann to Fermi-Dirac distribution. Under such condition, the quantum degeneracy effects start playing a significant role as the thermal de-Broglie wavelength for the plasma electrons becomes comparable or equal to that of average inter-electronic distance [18]. Quantum plasma's are generally made of electrons and ions or holes. The studies on quantum plasma got importance in several branches of applied physics specially in nano-science [19-21] as well as in laboratory plasma experiments [22-25] and in astrophysical scenario [26-28].\\
Pine [29] have treated an arbitrary collision-less quantum plasma environment as a dielectric medium and  derived the analytic form of dielectric function using Random Phase Approximation (RPA) method. Using such dielectric function, Sukla \textit{et. al.} [30] showed that the effective potential felt by a slowly moving test charge has two component: the usual near field Debye-Huckel screening term and far-field wake potential. Far field wake potential decays as the inverse cube of the distance between the origin of the test charge and the location of the observer. It is interesting to note that for far field, the effective potential of a moving ``\textit{test charge}'' in an isotropic collision-less classical plasma also falls off as the inverse cube of the distance of the observer from the test charge [31]. The effect of far field wake potential is very small on the binding energy of atom. Thus it is very much important to study devoted to the effect of near field wake potential on the binding energy of atom/ion. The only attempt in this context was made by Hu \textit{et. al.} [32]. They [32] have found that the near field wake potential is proportional to $\frac{1}{r^{2}}$ and $\cos\theta$; $r$ being the radial distance between the moving ion and the observer while $\theta$ is the angle between the radial vector and velocity vector of the ion. They [32] have used Meijer's G function in deriving the analytic form of the near field wake potential, where this G function violets the condition used in its definition [33]. This results in some anomalous findings in binding energy calculations \textit{e.g.} variationally over-bound energy levels \textit{w.r.t.} the energy levels of the free atom and the removal of degeneracy of the energy levels \textit{w.r.t.} the magnetic quantum number `$m$'. Even if we assume that their form of the potential to be correct, the energy levels should be Stark-like shifted due to `$\cos\theta$' term in the potential and due to obvious reason there is no possibility of getting Zeeman-like splitting without any perturbation \textit{e.g.} magnetic field which breaks the azimuthal symmetry of the the system.\\
To examine the influence of near field wake potential on the structural properties of moving atom/ion in quantum plasma, the analytic form of the potential has been derived in the present work using correct form of Meijer's G function [33] and its identities. The present derived potential is proportional to $rK_{0}(\frac{r}{\lambda_{q}})$ [$K_{0}(x)$ being the zeroth order Mac-Donald function or Modified Bessel's function of second kind [34] and $\lambda_{q}$ is the Debye parameter] and $\cos\theta$. Subsequently we have applied Rayleigh-Ritz variation method to obtain the binding energies of all states lying between $1s$ and $4f$ configuration of hydrogen-like carbon ion moving through Electron-Hole-Droplet (EHD) quantum plasma. In contrast to the findings of Hu \textit{et. al.} [32], no overbound result has been observed. Moreover the splitting of energy levels \textit{w.r.t.} $\lvert m\rvert$, has been observed which is purely Stark-like shifting due to oscillatory term in the potential. The details of the present methodology are given in section II, followed by results and discussion in section III and finally the conclusion is given in section IV.  
\section{\label{sec:level2}Method}
\subsection{Near-field potential felt by a slowly moving ``\textit{test charge}'' in quantum plasma:}
The field of a charge $q$ moving with a velocity $\vv{v}$ in a dielectric medium is given by the equation [18]
\begin{eqnarray}
\vv{\nabla}. \vv{D} =4\pi q \delta(\vv{r}-\vv{v}t)
\end{eqnarray}
Considering the quantum plasma environment as a linear dielectric medium, we have the relations $\vv{D}=\epsilon \vv{E}$; where the electric field $\vv{E}$ is derived from the scalar potential $\varphi$ as $\vv{E}=-\vv{\nabla}\varphi$. Equation (1) will be modified as
\begin{eqnarray}
-\vv{\nabla}\epsilon . \vv{\nabla}\varphi - \epsilon\nabla^{2}\varphi = 4\pi q \delta(\vv{r}-\vv{v}t)
\end{eqnarray}
\begin{figure}[htbp]
\includegraphics[width=0.5\textwidth]{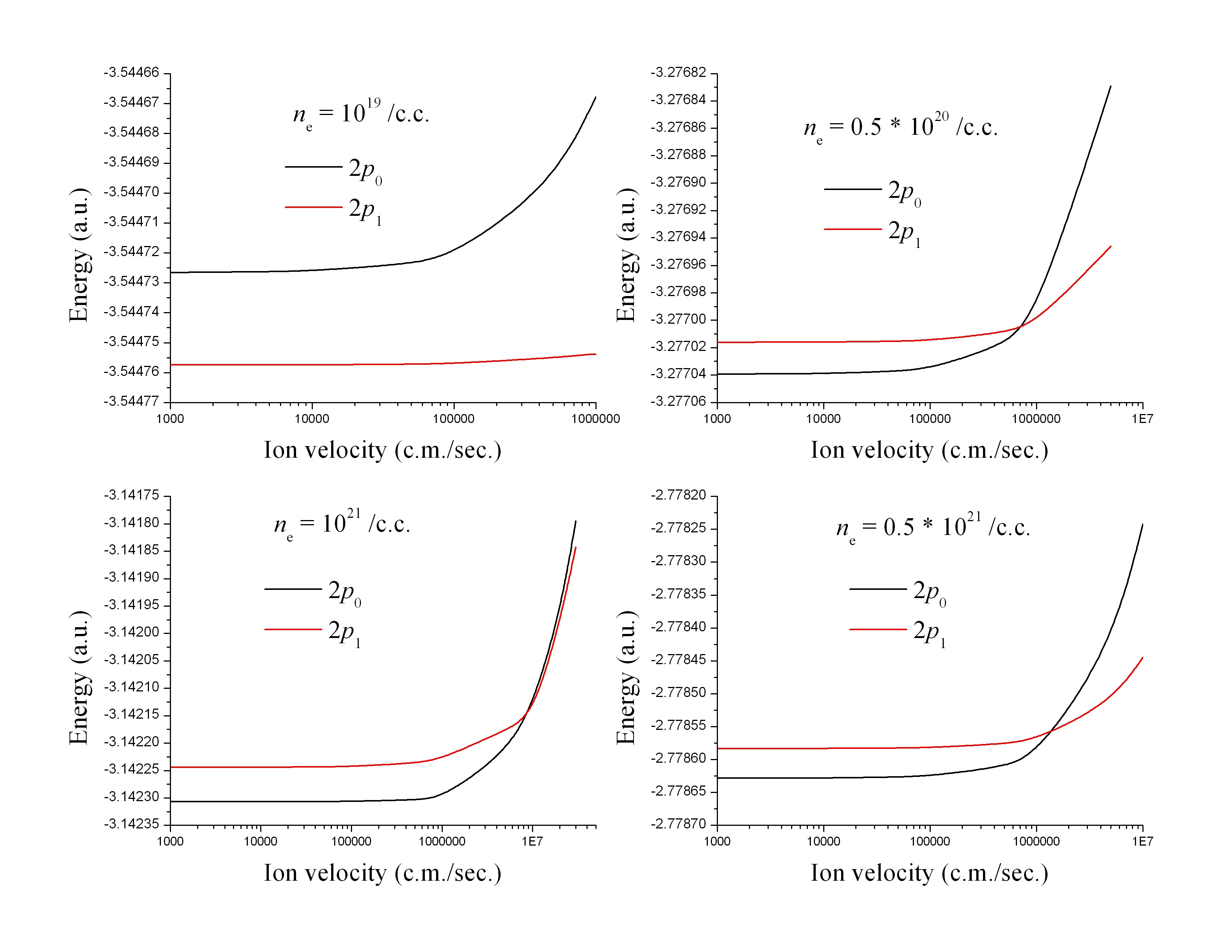}
\caption{Plot of energy values (in a.u.) of $2p_{0}$ and $2p_{1}$ states of C$^{5+}$ against ion velocity (in c.m./sec.) for different plasma electron densities (/c.c).}
\end{figure}
\begin{figure}[htbp]
\includegraphics[width=0.5\textwidth]{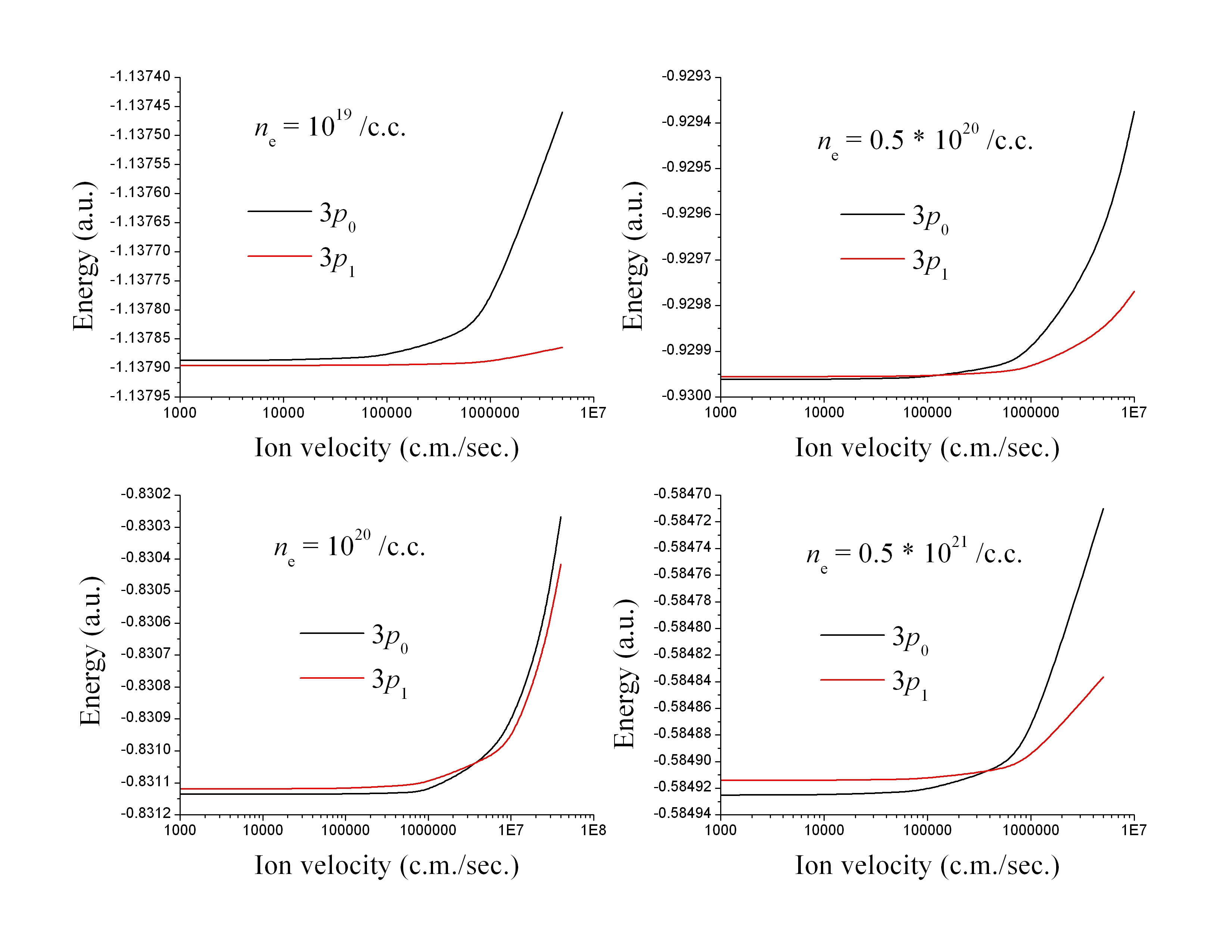}
\caption{Plot of energy values (in a.u.) of $3p_{0}$ and $3p_{1}$ states of C$^{5+}$ against ion velocity (in c.m./sec.) for different plasma electron densities (/c.c).}
\end{figure}
Taking Fourier transform on both sides of equation (2), we can write
\begin{eqnarray}
\varphi (k)=\frac{4\pi q}{(2\pi)^{\frac{3}{2}}}\frac{e^{-i\vv{k}.\vv{v}t}}{k^{2}\epsilon(k)}
\end{eqnarray}
The potential $\varphi (r)$ can be obtained by inverse Fourier transform [18] of equation (3) and given by,
\begin{eqnarray}
\varphi (\vv{r})=\frac{q}{2\pi^{2}}\int \frac{e^{i\vv{k}.\vv{r}}}{k^{2}\epsilon(\vv{k},\omega)}d^{3}\vv{k}
\end{eqnarray}
The dielectric function $\epsilon (k, \omega)$ for low frequency perturbation ($\omega \ll k v_{ts}$) is derived by Pines [29] as 
\begin{eqnarray}
\epsilon (\vv{k},\omega)=1+\sum_{s=e,h} \frac{K_{Fs}^{2}}{k^{2}}\left(1+i\frac{\pi}{2}\frac{\omega}{kv_{ts}}\right)
\end{eqnarray}
where, $v_{ts}=\frac{\hbar}{m_{s}}(3\pi^{2}n_{s})^{\frac{1}{3}}$ is the thermal velocity. The subscript $s$ used in the expression for thermal velocity ($v_{ts}$) means the species of the plasma. For electron-hole droplet plasma, the species means either electron ($e$) or hole ($h$); $m_{s}$ and $n_{s}$ are the effective mass and density respectively of the species $s$. For the present calculation $m_{h}=0.39M_{e}$ and $m_{e}=0.26M_{e}$ [35, 36] is taken; where $M_{e}$ is the rest mass of the electron. The Fermi-Thomas screening wave number $K_{Fs}$ is defined as $K_{Fs}=\frac{\sqrt{3}\omega_{ps}}{v_{ts}}$ where the plasma oscillation frequency $\omega_{ps}=\left(\frac{4\pi n_{s}e^{2}}{m_{s}}\right)^{\frac{1}{2}}$. It should be mentioned that the Debye length $\lambda_{s}=\frac{1}{K_{Fs}}$. Equation (5) can be rearranged as,
\begin{eqnarray}
\epsilon (\vv{k},\omega)&=&\frac{1+k^{2}\lambda_{q}^{2}}{k^{2}\lambda_{q}^{2}}\left[1+i\frac{\pi}{2}\frac{\omega\lambda_{q}^{2}}{k(1+k^{2}\lambda_{q}^{2})}\sum \frac{1}{v_{ts}\lambda_{s}^{2}} \right]
\end{eqnarray}
where $\frac{1}{\lambda_{q}^{2}}=\sum_{s=e,h}\frac{1}{\lambda_{s}^{2}}$. Pines [29] obtained equation (5) after performing complex integration where the pole position is at $\omega =-\vv{k}.\vv{v}$ [30, 37]. The velocity ($v$) of the ions are chosen so that the thermal Mac number [37] remains below unity. For $v <v_{ts}$, we can get
\begin{eqnarray}
\frac{1}{\epsilon(\vv{k},\omega)}&\approx& \frac{k^{2}\lambda_{q}^{2}}{1+k^{2}\lambda_{q}^{2}} + i\frac{\pi}{2}\frac{k\lambda_{q}^{4}}{(1+k^{2}\lambda_{q}^{2})^{2}}\vv{k}.\vv{v}\nonumber\\
&\times&\sum_{s=e,h}\frac{1}{v_{ts}\lambda_{s}^{2}}
\end{eqnarray}
combining equation (4) and (7), we obtain 
\begin{eqnarray}
\varphi =\varphi_{1}+\varphi_{2}
\end{eqnarray} 
where
\begin{eqnarray}
\varphi_{1}=\frac{q}{2\pi^{2}}\int \frac{\lambda_{q}^{2}}{1+k^{2}\lambda_{q}^{2}}e^{i\vv{k}.\vv{r}}d^{3}\vv{k}
\end{eqnarray}
\begin{figure}[htbp]
\includegraphics[width=0.5\textwidth]{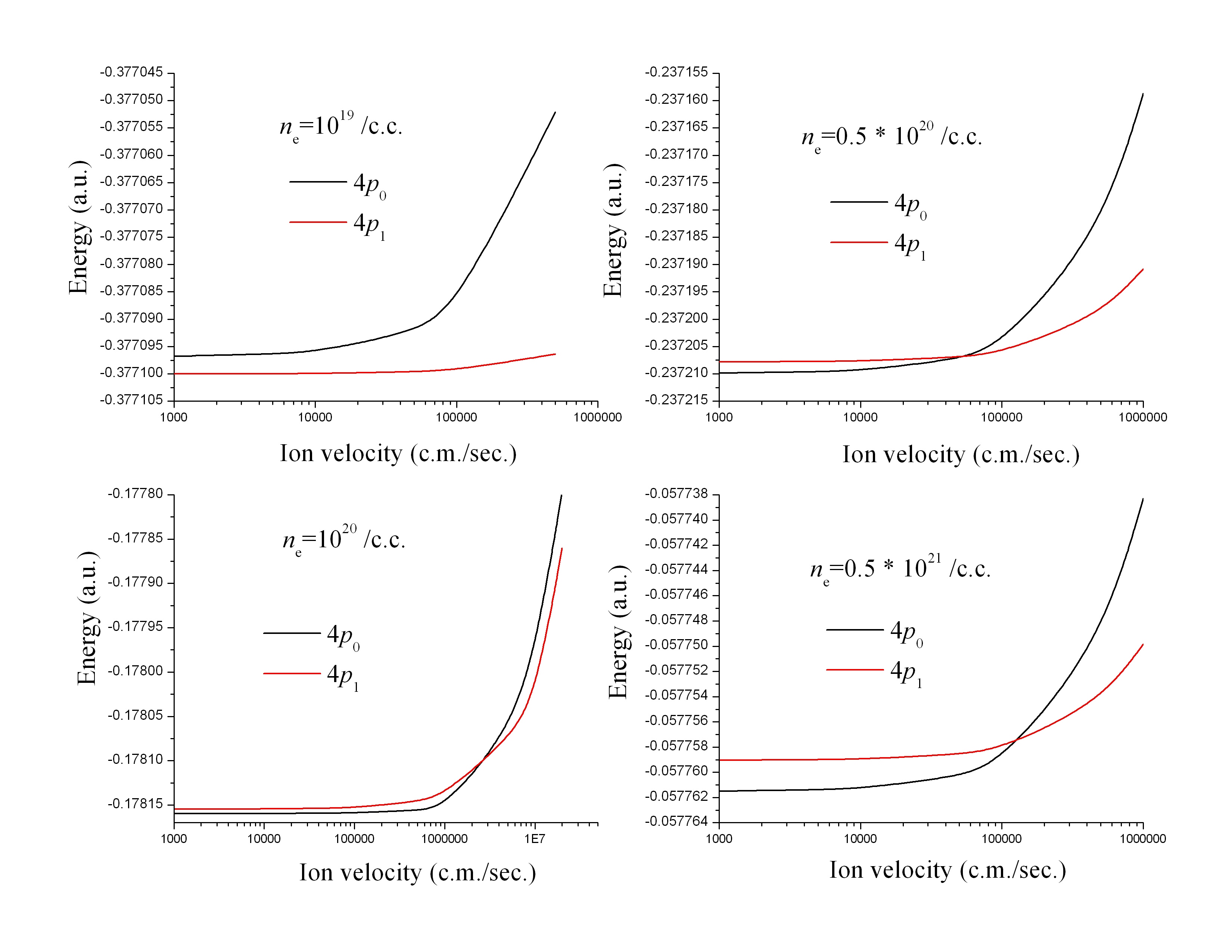}
\caption{Plot of energy values (in a.u.) of $4p_{0}$ and $4p_{1}$ states of C$^{5+}$ against ion velocity (in c.m./sec.) for different plasma electron densities (/c.c).}
\end{figure}
\begin{figure}[htbp]
\includegraphics[width=0.5\textwidth]{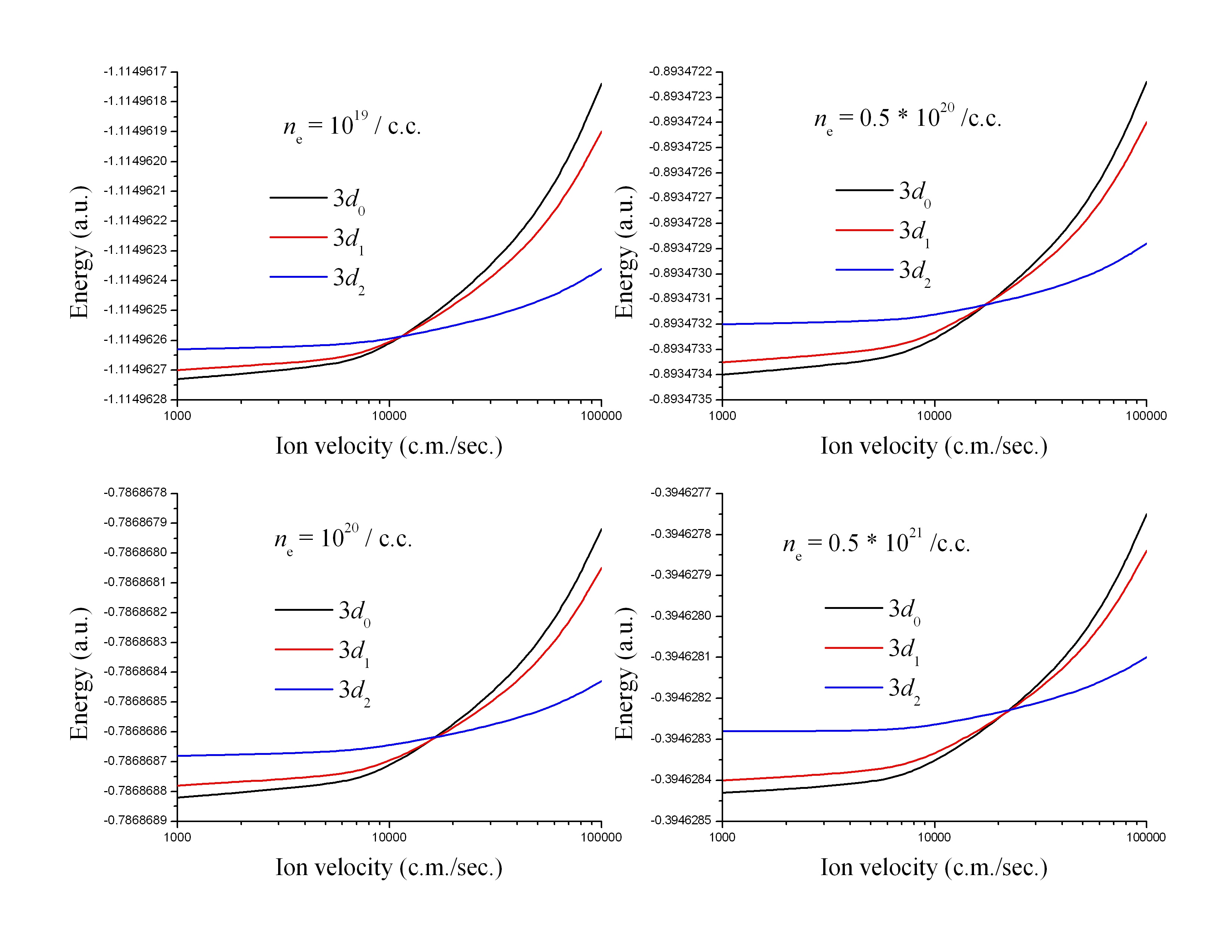}
\caption{Plot of energy values (in a.u.) of $3d_{0}$, $3d_{1}$ and $3d_{2}$ states of C$^{5+}$ against ion velocity (in c.m./sec.) for different plasma electron densities (/c.c).}
\end{figure}
In the spherical polar coordinate ($k$, $\sigma$, $\tau$) system, the volume element is given as $d^{3}\vv{k}=k^{2} \sin\sigma d\sigma d\tau dk$. Integrating over $\sigma$ and $\tau$, equation (9) reduces to
\begin{eqnarray}
\varphi_{1}=\frac{2q\lambda_{q}^{2}}{\pi r}\int_{o}^{\infty} \frac{k}{1+k^{2}\lambda_{q}^{2}}\sin kr dk = \frac{q}{r}e^{-\frac{r}{\lambda_{q}}}
\end{eqnarray} 
$\varphi_{1}$ is the well-known Debye-Huckel screening potential [4, 16].\\
The second term of equation (8) is given by
\begin{eqnarray}
\varphi_{2}&=&i\frac{\pi}{2}\frac{q}{2\pi^{2}}\int \frac{\lambda_{q}^{4}}{k(1+k^{2}\lambda_{q}^{2})^{2}}\vv{k}.\vv{v}\sum_{s=e,h}\frac{1}{v_{ts}\lambda_{s}^{2}}\nonumber\\
&\times&e^{i\vv{k}.\vv{r}}d^{3}\vv{k}
\end{eqnarray} 
Performing integration over the azimuthal angle,  equation (11) reduces to 
\begin{eqnarray}
\varphi_{2}&=&\frac{i}{2}qv\lambda_{q}^{4}\sum_{s=e,h}\frac{1}{v_{ts}\lambda_{s}^{2}}\int_{o}^{\infty}\frac{k^{2}}{(1+k^{2}\lambda_{q}^{2})^{2}}\nonumber\\
&\times&\int_{0}^{\pi}\cos(\sigma +\theta)e^{ikr\cos\sigma}\sin\sigma d\sigma
\end{eqnarray}
where, $\theta$ is the angle between $\vv{r}$ and $\vv{v}$, ($\theta + \sigma$) is the angle between $\vv{k}$ and $\vv{v}$. The polar angle part of the integral can be written as
\begin{eqnarray}
\int_{0}^{\pi}\cos(\sigma +\theta)e^{ikr\cos\sigma}\sin\sigma d\sigma~~~~~~~~~~~~~~~~~~~~~ \nonumber\\
=\cos\theta .I_{1}-\sin\theta . I_{2}
\end{eqnarray}
where
\begin{eqnarray}
I_{1}=\int_{0}^{\pi}\cos\sigma e^{ikr\cos\sigma}\sin\sigma d\sigma =-\frac{2}{i}j_{1}(kr)
\end{eqnarray}
and
\begin{eqnarray}
I_{2}=\int_{0}^{\pi}\sin\sigma e^{ikr\cos\sigma}\sin\sigma d\sigma 
= \frac{\pi}{2}\left[j_{0}(kr)+j_{2}(kr)\right]\nonumber\\
\end{eqnarray}
where $j_{l}(x)$ is the spherical Bessel function of first kind. Neglecting the imaginary part, equation (12) turns to
\begin{eqnarray}
\varphi_{2}=-qv\lambda_{q}^{4}\sum_{s=e,h}\frac{1}{v_{ts}\lambda_{s}^{2}}\cos\theta \int_{0}^{\infty}\frac{k^{2}j_{1}(kr)}{(1+k^{2}\lambda_{q}^{2})^{2}}dk
\end{eqnarray}
Using the following two identities [33] 
\begin{eqnarray}
j_{\nu}(z)=G_{0~2}^{1~0}\left(\begin{matrix}
-\\
\frac{\nu}{2}, \frac{\nu}{2}
\end{matrix} \vert \frac{z^{2}}{4} \right)
\end{eqnarray}
and [38]
\begin{eqnarray}
\frac{z^{\beta}}{(1+az^{b})^{\alpha}}=\frac{a^{-\frac{\beta}{b}}}{\Gamma(\alpha)}G_{1~1}^{1~1}\left(\begin{matrix}
1-\alpha +\frac{\beta}{b}\\
\frac{\beta}{b}
\end{matrix} \vert az^{b} \right)
\end{eqnarray}
equation (16) can be written as
\begin{eqnarray}
\varphi_{2}&=&-\frac{qv\lambda_{q}^{3}}{2}\sum_{s=e,h}\frac{1}{v_{ts}\lambda_{s}^{2}}\cos\theta \int_{0}^{\infty}G_{1~1}^{1~1}\left(\begin{matrix}
-\frac{1}{2}\\
\frac{1}{2}
\end{matrix} \vert \lambda_{q}^{2}k^{2} \right)\nonumber\\
&\times&G_{0~2}^{1~0}\left(\begin{matrix}
-\\
\frac{1}{2}, -\frac{1}{2}
\end{matrix} \vert \frac{k^{2}r^{2}}{4} \right)d(k^{2})
\end{eqnarray}
where $G_{p~q}^{m~n}\left(\begin{matrix}
a_{1},...,a_{p}\\
b_{1},...,b_{q}
\end{matrix} \vert x \right)$ is the Meijer's $G$ function [33] defined as
\begin{eqnarray}
G_{p~q}^{m~n}\left(\begin{matrix}
a_{1},...,a_{p}\\
b_{1},...,b_{q}
\end{matrix} \vert x \right)=~~~~~~~~~~~~~~~~~~~~~~~~~~~~~~~\nonumber\\
\frac{1}{2\pi i}\int \frac{\prod_{j=1}^{m}\Gamma (b_{j}-s)\prod_{j=1}^{n}\Gamma (1-a_{j}+s)}{\prod_{j=m+1}^{q}\Gamma (1-b_{j}+s)\prod_{j=n+1}^{p}\Gamma (a_{j}-s)}\nonumber\\
\times x^{s}ds
\end{eqnarray}
with the constraints $0\leq m\leq q$ and $0\leq n\leq p$ and $\Gamma (n)$ is the Euler Gamma function. 
Using the identities [33] given below
\begin{eqnarray}
\int_{0}^{\infty}G_{u~v}^{s~t}\left(\begin{matrix}
c_{1},...,c_{u}\\
d_{1},...,d_{v}
\end{matrix} \vert \xi x \right)G_{p~q}^{m~n}\left(\begin{matrix}
a_{1},...,a_{p}\\
b_{1},...,b_{q}
\end{matrix} \vert \eta x \right)dx ~~~~~~~~~\nonumber\\ 
=\frac{1}{\xi}G_{p+v~q+u}^{t+m~s+n}\left(\begin{matrix}
a_{1},...,a_{n},-d_{1},...,-d_{v},a_{n+1},...,a_{p}\\
b_{1},...,b_{m},-c_{1},...,-c_{u},b_{m+1},...,b_{q}
\end{matrix} \vert \frac{\eta}{\xi} \right)\nonumber\\
\end{eqnarray}
\begin{eqnarray}
x^{k}G_{p~q}^{m~n}\left(\begin{matrix}
a_{1},...,a_{p}\\
b_{1},...,b_{q}
\end{matrix} \vert x \right)=G_{p~q}^{m~n}\left(\begin{matrix}
a_{1}+k,...,a_{p}+k\\
b_{1}+k,...,b_{q}+k
\end{matrix} \vert  x \right)\nonumber\\
\end{eqnarray}
\begin{eqnarray}
G_{p~q}^{m~n}\left(\begin{matrix}
a_{1},...,a_{p}\\
b_{1},...,b_{q-1},a_{1}
\end{matrix} \vert  x \right)=G_{p-1~q-1}^{m~n-1}\left(\begin{matrix}
a_{2},...,a_{p}\\
b_{1},...,b_{q}-1
\end{matrix} \vert  x \right)\nonumber\\
\end{eqnarray}
equation (19) modifies as,
\begin{eqnarray}
\varphi_{2} = -\frac{qv\lambda_{q}^{2}}{r}\sum_{s=e,h}\frac{1}{v_{ts}\lambda_{s}^{2}}G_{0~2}^{2~0}\left(\begin{matrix}
-\\
1,1
\end{matrix} \vert \frac{r^{2}}{4\lambda_{q}^{2}} \right)\cos\theta
\end{eqnarray}
It is interesting to note that, the Meijer's $G$ function [33] appears in the above equation converges iff the argument \textit{i.e.} $\frac{r^{2}}{4\lambda_{q}^{2}}$ becomes less than unity \textit{i.e.} $r<2\lambda_{q}$. To get the final form of the potential $\varphi_{2}$, we have used the identity [33] as given by
\begin{eqnarray}
2^{\mu -1}G_{0~2}^{2~0}\left(\begin{matrix}
-\\
\frac{\mu}{2}+\frac{\nu}{2},\frac{\mu}{2}-\frac{\nu}{2}
\end{matrix} \vert x \right)=x^{\mu}K_{\nu}(x)
\end{eqnarray}
The final form of near-field wake potential $\varphi_{2}$ is given by
\begin{eqnarray}
\varphi_{2}=-\frac{qv}{2} \sum_{s=e,h}\frac{1}{v_{ts}\lambda_{s}^{2}} r K_{0}\left(\frac{r}{\lambda_{q}}\right)\cos\theta
\end{eqnarray}
where $K_{\nu}(x)$ is the Mac-donald function or modified Bessel function of second kind. It is interesting to note that, similar kind of radial dependence of the potential was obtained by Frolov [39] in case of short range interaction between two point electric charges.
\subsection{Structure calculation of slowly moving hydrogen-like ion in quantum plasma:}
The modified non-relativistic Hamiltonian of a slowly moving hydrogen-like ion in the presence of an external quantum plasma environment can be represented by
\begin{eqnarray}
H=-\frac{1}{2}\nabla^{2}+V_{eff}(r,\theta)
\end{eqnarray}
where the near-field effective potential $V_{eff}(r,\theta)$ composed of two parts as
\begin{eqnarray}
V_{eff}(r,\theta)=V_{d}(r)+V_{w}(r,\theta)
\end{eqnarray}
here, $V_{d}(r)$ is the Debye-Huckel screening potential given by
\begin{eqnarray}
V_{d}(r)=-\frac{Z}{r}e^{-\mu r}
\end{eqnarray}
where, $Z$ is the atomic number of the moving ion and $\mu$ is the Debye screening parameter related to Debye length as $\mu=\frac{1}{\lambda_{q}}$.\\
\begin{sidewaystable}[htbp]
\begin{center}
\caption {\rm {Convergence of the energy eigenvalues ($a.u.$) of $1s_{0}$, $2p_{1}$, $3d_{2}$ and $4f_{3}$ states of C$^{5+}$ moving in quantum plasma. The number density ($n_{e}$) of electrons is taken as $10^{19}$/c.c. while the speed of the ion ($v$) is $10^{3}$ c.m./sec. $N$ represents the total number of terms in the basis set.}}
\begin{scriptsize}
\begin{tabular}{c c c c c c c c c c c}\\
\hline\hline\vspace{-.15cm}\\
$N$&$1s_{0}$&&$N$&$2p_{1}$&&$N$&$3d_{2}$&&$N$&$4f_{3}$\\
\cline{1-2}\cline{4-5}\cline{7-8}\cline{10-11}
\vspace{-.15cm}\\
1&16.996 989 379&& 2 &3.516 726 519&&3&1.101 328 256&&4&0.328 549 543\\	
3&16.996 997 305&& 5 &3.544 757 189&&7&1.114 962 499&&9&0.329 492 549\\
6&16.996 997 518&& 9 &3.544 757 475&&12&1.114 962 616&&15&0.329 493 091\\
15&16.996 997 543&&20&3.544 757 476&&25&1.114 962 632&&22&0.329 493 108\\
28&16.996 997 543&&35&3.544 757 476&&42&1.114 962 632&&39&0.329 493 108\\
45&16.996 997 543&&54&3.544 757 476&&52&1.114 962 632&&49&0.329 493 108\\
\hline\hline
\end{tabular}
\end{scriptsize}
\end{center}
\end{sidewaystable}
The near-field wake potential $V_{w}$ ($r$, $\theta$) is given by
\begin{eqnarray}
V_{w}(r,\theta)=\zeta r K_{0}\left(\frac{r}{\lambda_{q}}\right)\cos\theta
\end{eqnarray}
where the wake field coefficient is defined as $\zeta = \frac{Zv}{2} \sum_{s=e,h}\frac{1}{v_{ts}\lambda_{s}^{2}}$.\\
The variational equation for any arbitrary angular momentum state of one electron system is given by
\begin{eqnarray}
\delta\int\left[\left(\frac{\partial\Psi}{\partial r}\right)^{2}+\frac{1}{r^{2}}\left(\frac{\partial\Psi}{\partial \theta}\right)^{2}+~~~~~~~~~~~~~~~~~~~~~\right.\nonumber\\
\left.\frac{1}{r^{2}\sin^{2}\theta}\left(\frac{\partial\Psi}{\partial \phi}\right)^{2}+2(V_{eff}-E)\Psi^{2}\right]dv_{r,\theta,\phi}=0
\end{eqnarray}
subject to the normalization condition
\begin{eqnarray}
\int \Psi^{2} dv_{r,\theta ,\phi}=1
\end{eqnarray}
The trial wavefunction is taken as,
\begin{eqnarray}
\Psi (r,\theta ,\phi)=f(r)A_{l,m}(\theta, \phi)
\end{eqnarray}
where, the radial part is given by
\begin{eqnarray}
f(r)=\sum_{i=1}^{N}C_{i}\chi_{i}(r)
\end{eqnarray}  
with $\chi_{i}(r)=r^{n_{i}}e^{-\alpha_{i}r}$ and the trial angular part is given by
\begin{eqnarray}
A_{lm}(\theta , \phi)=(\gamma +\beta \cos\theta)Y_{lm}(\theta , \phi)
\end{eqnarray}
where $Y_{lm}(\theta , \phi)$ being the spherical harmonics.\\
In order to calculate the matrix elements of the Hamiltonian we have used the following integral [33] as 
\begin{eqnarray}
\int_{0}^{\infty}x^{\mu -1}e^{-\alpha x}K_{\nu}(\beta x)dx~~~~~~~~~~~~~~~~~~~~~~~~~~~~\nonumber\\
=\frac{\sqrt{\pi}(2\beta)^{\nu}}{(\alpha +\beta)^{\mu +\nu}}\frac{\Gamma (\mu +\nu)\Gamma (\mu -\nu)}{\Gamma (\mu -\frac{1}{2})}~~~~~~~~~~~\nonumber\\
~~~~\times F\left(\mu +\nu , \nu +\frac{1}{2}, \mu +\frac{1}{2};\frac{\alpha -\beta}{\alpha +\beta}\right)
\end{eqnarray}
where, $F$ is the confluent Hypergeometric function and Re $\mu >$ $\vert$ Re $\nu$ $\vert$ and Re $(\alpha +\beta)>0$.\\
Finally we have solved the generalized eigenvalue equation [40] given as,
\begin{eqnarray}
\underline{\underline{H}}~\underline{C}=E\underline{\underline{S}}~\underline{C}
\end{eqnarray}
where $\underline{\underline{H}}$ is the Hamiltonian matrix, $\underline{\underline{S}}$ is the overlap matrix and $E$'s are the energy eigenroots. The non-linear parameters $\alpha_{i}$'s, $\beta$ and $\gamma$ are being optimized by using Nelder-Mead procedure [41]. The convergence behavior of the energy
eigenvalues has been checked by increasing the number of terms in the wave function to ensure the accuracy of the present method. All calculations are being carried out in quadruple precision. Atomic units are used throughout.
\section{\label{sec:level2}Results and Discussions}
\begin{table}[htbp]
\begin{center}
\caption {\rm {The energy eigenvalues -$E$ (a.u.) of $1s_{0}$ states of C$^{5+}$ moving in quantum plasma having different set of electron number density ($n_{e}$/c.c) and ion velocity ($v$ c.m./sec).}}
\begin{scriptsize}
\begin{tabular}{c c c}\\
\hline\hline\vspace{-.2cm}\\
&&-$E_{1s}$ (a.u.)\\
\cline{3-3}\vspace{-.2cm}\\
$n_{e}$ (/c.c.)&$v$ (c.m./sec.)&$\lvert m\rvert$=0\\
\hline\vspace{-.2cm}\\
0        &	     0&	18.00000000\\
$10^{19}$&	     0&	16.99701207\\
	     &$10^{3}$&	16.99699754\\
	     &$10^{5}$&	16.99699738\\
	     &$10^{7}$&	16.99698206\\
$10^{20}$&       0&	16.54217234\\
	     &$10^{3}$&	16.54214839\\
	     &$10^{5}$&	16.54214822\\
	     &$10^{7}$&	16.54213093\\
$10^{21}$&	     0&	15.89053458\\
	     &$10^{3}$&	15.89046598\\
	     &$10^{5}$&	15.89046573\\
	     &$10^{7}$&	15.89044045\\
$10^{22}$&	     0&	14.96731440\\
	     &$10^{3}$&	14.96727643\\
	     &$10^{5}$&	14.96727627\\
	     &$10^{7}$&	14.96726037\\
$10^{23}$&	     0&	13.68055489\\
	     &$10^{3}$&	13.68051129\\
	     &$10^{5}$&	13.68051115\\
	     &$10^{7}$&	13.68049712\\
\hline\hline
\end{tabular}
\end{scriptsize}
\end{center}
\end{table}
\begin{table}[htbp]
\begin{center}
\caption {\rm {The energy eigenvalues -$E$ (a.u.) of $2s_{0}$, $2p_{0}$ and $2p_{1}$ states of C$^{5+}$ moving in quantum plasma having different set of electron number density ($n_{e}$/c.c) and ion velocity ($v$ c.m./sec).}}
\begin{scriptsize}
\begin{tabular}{c c c c c c}\\
\hline\hline\vspace{-.2cm}\\
&&-$E_{2s}$ (a.u.)&&\multicolumn{2}{c}{-$E_{2p}$ (a.u.)}\\
\cline{3-3} \cline{5-6}\vspace{-.2cm}\\
$n_{e}$ (/c.c.)&$v$ (c.m./sec.)&$\lvert m\rvert$=0&&$\lvert m\rvert$=0&$\lvert m\rvert$=1 \\
\hline\vspace{-.2cm}\\
0        &     	 0&	4.50000000&&	4.50000000&	4.50000000\\
$10^{19}$&	     0&	3.55780856&&	3.54476161&	3.54476161\\
	     &$10^{3}$&	3.55780693&&	3.54472655&	3.54475747\\
	     &$10^{5}$&	3.55780657&&	3.54472073&	3.54475700\\
	     &$10^{7}$&	3.55777061&&	3.54418783&	3.54449515\\
$10^{20}$&	     0&	3.16902746&&	3.14230776&	3.14230777\\
	     &$10^{3}$&	3.16902456&&	3.14229037&	3.14224439\\
	     &$10^{5}$&	3.16902418&&	3.14229621&	3.14224287\\
	     &$10^{7}$&	3.16898702&&	3.14223675&	3.14215211\\
$10^{21}$&	     0&	2.65209786&&	2.59865596&	2.59865596\\
	     &$10^{3}$&	2.65208981&&	2.59862497&	2.59858157\\
	     &$10^{5}$&	2.65208933&&	2.59862135&	2.59858025\\
	     &$10^{7}$&	2.65204099&&	2.59825988&	2.59844818\\
$10^{22}$&	     0&	1.99849293&&	1.89549274&	1.89549274\\
	     &$10^{3}$&	1.99848876&&	1.89542088&	1.89549069\\
	     &$10^{5}$&	1.99848850&&	1.89541666&	1.89549053\\
	     &$10^{7}$&	1.99846268&&	1.89483600&	1.89547373\\
$10^{23}$&	     0&	1.23890690&&	1.05303897&	1.05303897\\
	     &$10^{3}$&	1.23890275&&	1.05281193&	1.05303710\\
	     &$10^{5}$&	1.23890257&&	1.05280651&	1.05303730\\
	     &$10^{7}$&	1.23888478&&	1.05229719&	1.05302671\\
 \hline\hline
\end{tabular}
\end{scriptsize}
\end{center}
\end{table}
We have calculated the energy eigenvalues of $ns_{0}$ [the principal quantum number, $n=1-4$ and the subscript denotes the values of the azimuthal quantum number]; $np_{0}$, $np_{1}$ [$n=2-4$]; $nd_{0}$, $nd_{1}$, $nd_{2}$ [$n=3-4$] and $nf_{0}$, $nf_{1}$, $nf_{2}$, $nf_{3}$  [$n=4$] states of $C^{5+}$ ion. The plasma electron densities ($n_{e}$) are chosen in the range $10^{19}-10^{23}$ /c.c. while for each value of plasma density ($n_{e}$), the ion velocities ($v$) are ranging from $10^{3}-10^{7}$ c.m./sec. Table-1 displays the results for convergence of energy eigenvalues for $1s_{0}$, $2p_{1}$, $3d_{2}$ and $4f_{3}$ states with plasma density ($n_{e}$) $10^{19}$/c.c. and ion velocity ($v$) $10^{3}$ c.m./sec. It is evident from table-1 that the energy eigenvalues converge upto 9th decimal place in each cases. Similar convergence of energy values are being obtained for all the calculations done in the present communication.\\
\begin{figure}[htbp]
\includegraphics[width=0.5\textwidth]{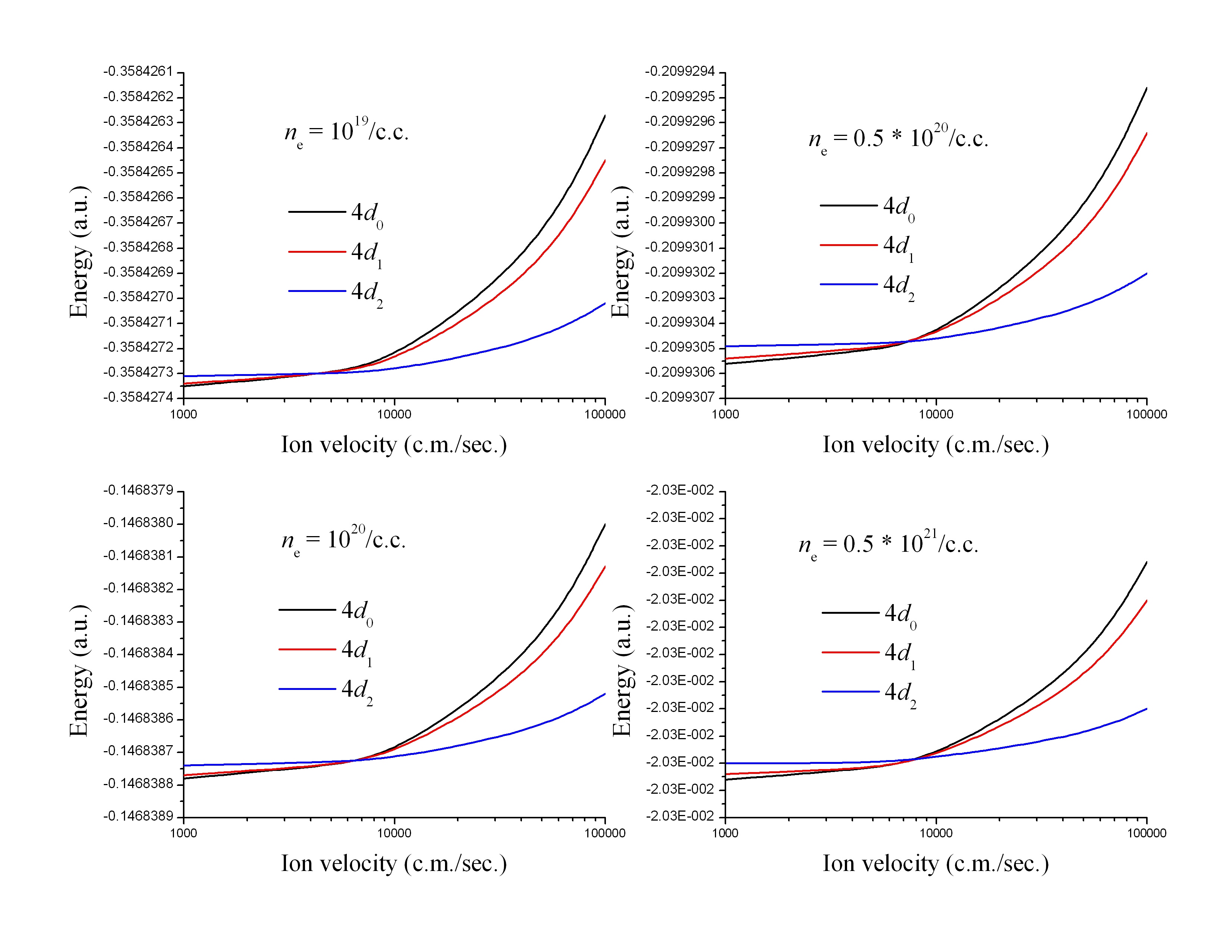}
\caption{Plot of energy values (in a.u.) of $4d_{0}$, $4d_{1}$ and $4d_{2}$ states of C$^{5+}$ against ion velocity (in c.m./sec.) for different plasma electron densities (/c.c).}
\end{figure}
\begin{figure}[htbp]
\includegraphics[width=0.5\textwidth]{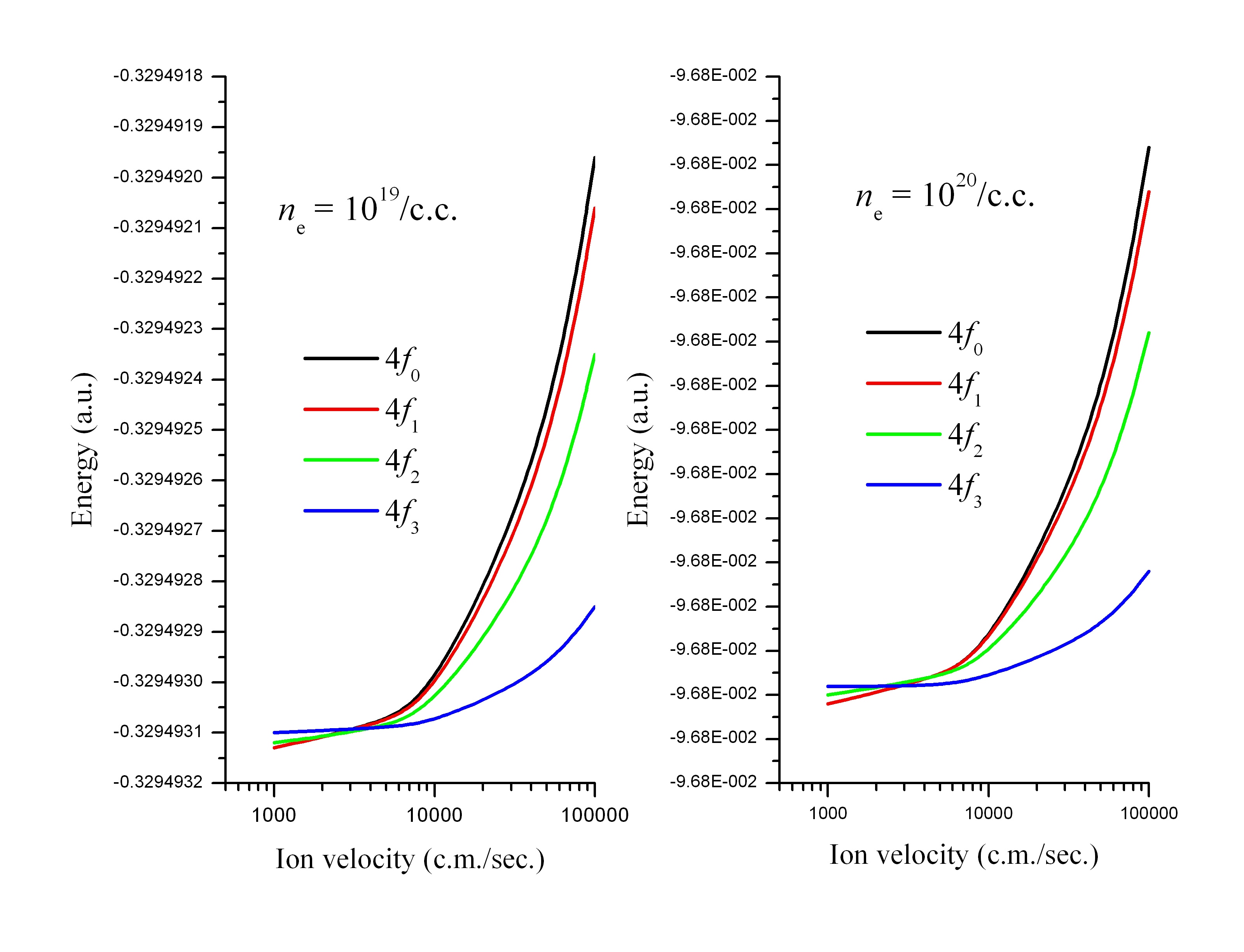}
\caption{Plot of energy values (in a.u.) of $4f_{0}$, $4f_{1}$, $4f_{2}$ and $4f_{3}$ states of C$^{5+}$ against ion velocity (in c.m./sec.) for different plasma electron densities (/c.c).}
\end{figure}
\begin{sidewaystable}[htbp]
\begin{center}
\caption {\rm {The energy eigenvalues -$E$ (a.u.) of $3s_{0}$, $3p_{0}$, $3p_{1}$, $3d_{0}$, $3d_{1}$ and $3d_{2}$ states of C$^{5+}$ moving in quantum plasmas having different set of electron number density ($n_{e}$/c.c) and ion velocity ($v$ c.m./sec).}}
\begin{scriptsize}
\begin{tabular}{c c c c c c c c c c}\\
\hline\hline\vspace{-.2cm}\\
&&-$E_{3s}$ (a.u.)&&\multicolumn{2}{c}{-$E_{3p}$ (a.u.)}&&\multicolumn{3}{c}{-$E_{3d}$ (a.u.)}\\
\cline{3-3} \cline{5-6}\cline{8-10}\vspace{-.2cm}\\
$n_{e}$ (/c.c.)&$v$ (c.m./sec.)&$\lvert m\rvert$=0&&$\lvert m\rvert$=0&$\lvert m\rvert$=1&&$\lvert m\rvert$=0&$\lvert m\rvert$=1&$\lvert m\rvert$=2 \\
\hline\vspace{-.2cm}\\
0        &	     0&	2.00000000&&	2.00000000&	2.00000000&&	2.00000000&	2.00000000&	2.00000000\\
$10^{19}$&	     0&	1.14923194&&	1.13789722&	1.13789723&&	1.11496287&	1.11496287&	1.11496287\\
	     &$10^{3}$&	1.14923145&&	1.13788733&	1.13789606&&	1.11496273&	1.11496270&	1.11496263\\
	     &$10^{5}$&	1.14923096&&	1.13787887&	1.13789537&&	1.11496174&	1.11496190&	1.11496236\\
	     &$10^{7}$&	1.14918232&&	1.13709896&	1.13759242&&	1.11486312&	1.11488121&	1.11493546\\
$10^{20}$&	     0&	0.85275345&&	0.83113577&	0.83113577&&	0.78686902&	0.78686902&	0.78686902\\
	     &$10^{3}$&	0.85275269&&	0.83113114&	0.83111901&&	0.78686882&	0.78686878&	0.78686868\\
	     &$10^{5}$&	0.85275224&&	0.83113021& 0.83111706&&	0.78686792& 0.78686805&	0.78686843\\
	     &$10^{7}$&	0.85270789&&	0.83104578&	0.83097475&&	0.78677385&	0.78679108&	0.78684275\\
$10^{21}$&	     0&	0.51307024&&	0.47479205&	0.47479205&&	0.39462864&	0.39462864&	0.39462864\\
	     &$10^{3}$&	0.51306840&&	0.47478483&	0.47477478&&	0.39462843&	0.39462840&	0.39462828\\
	     &$10^{5}$&	0.51306792&&	0.47478106&	0.47477341&&	0.39462775&	0.39462784&	0.39462810\\
	     &$10^{7}$&	0.51302066&&	0.47440449&	0.47463583&&	0.39455982&	0.39457226&	0.39460957\\
$10^{22}$&	     0&	0.18762946&&	0.13082887&	0.13082887&&		&0.01136017	&0.01136017\\
	     &$10^{3}$&	0.18762878&&	0.13081719&	0.13082853&&		&0.01136002	&0.01135996\\
	     &$10^{5}$&	0.18762861&&	0.13081432&	0.13082842&&		&0.01135976	&0.01135987\\
	     &$10^{7}$&	0.18761157&&	0.13044258&	0.13081701&&		&0.01133326	&0.01135103\\
$10^{23}$&	     0&	0.00358750&&              &&&&&\\					
	     &$10^{3}$&	0.00358738&&              &&&&&\\					
	     &$10^{5}$&	0.00358736&&              &&&&&\\					
	     &$10^{7}$&	0.00358535&&              &&&&&\\				
 \hline\hline
\end{tabular}
\end{scriptsize}
\end{center}
\end{sidewaystable}
\begin{sidewaystable}[htbp]
\begin{center}
\caption {\rm {The energy eigenvalues -$E$ (a.u.) of $4s_{0}$, $4p_{0}$, $4p_{1}$, $4d_{0}$, $4d_{1}$, $4d_{2}$, $4f_{0}$, $4f_{1}$, $4f_{2}$ and $4f_{3}$ states of C$^{5+}$ moving in quantum plasmas having different set of electron number density ($n_{e}$/c.c) and ion velocity ($v$ c.m./sec).}}
\begin{scriptsize}
\begin{tabular}{c c c c c c c c c c c c c c c}\\
\hline\hline\vspace{-.2cm}\\
&&-$E_{4s}$ (a.u.)&&\multicolumn{2}{c}{-$E_{4p}$ (a.u.)}&&\multicolumn{3}{c}{-$E_{4d}$ (a.u.)}&&\multicolumn{4}{c}{-$E_{4f}$ (a.u.)}\\
\cline{3-3} \cline{5-6}\cline{8-10} \cline{12-15}\vspace{-.2cm}\\
$n_{e}$ (/c.c.)&$v$ (c.m./sec.)&$\lvert m\rvert$=0&&$\lvert m\rvert$=0&$\lvert m\rvert$=1&&$\lvert m\rvert$=0&$\lvert m\rvert$=1&$\lvert m\rvert$=2&&$\lvert m\rvert$=0&$\lvert m\rvert$=1&$\lvert m\rvert$=2&$\lvert m\rvert$=3 \\
\hline\vspace{-.2cm}\\
0&	0&	1.12500000&&	1.12500000&	1.12500000&&	1.12500000&	1.12500000&	1.12500000&&	1.12500000&	1.12500000&	1.12500000&	1.12500000\\
$10^{19}$&	0&	0.38627150&&	0.37710044&	0.37710044&&	0.35842741&	0.35842741&	0.35842741&&	0.32949318&	0.32949318&	0.32949318&	0.32949318\\
	&$10^{3}$&	0.38627133&&	0.37709676&	0.37710000&&	0.35842735&	0.35842734&	0.35842731&&	0.32949313&	0.32949313&	0.32949312&	0.32949310\\
	&$10^{5}$&	0.38627083&&	0.37708789&	0.37709929&&	0.35842627&	0.35842645&	0.35842702&&	0.32949196&	0.32949206&	0.32949235&	0.32949285\\
	&$10^{7}$&	0.38622101&&	0.37626956&	0.37679442&&	0.35831828&	0.35833810&	0.35839757&&	0.32936955&	0.32938029&	0.32941613&	0.32946744\\
$10^{20}$&	0&	0.19334518&&	0.17815971&	0.17815971&&	0.14683884&	0.14683884&	0.14683884&&	0.09678684&	0.09678684&	0.09678684&	0.09678684\\
	&$10^{3}$&	0.19334495&&	0.17815827&	0.17815457&&	0.14683878&	0.14683877&	0.14683874&&	0.09678681&	0.09678681&	0.09678680&	0.09678679\\
	&$10^{5}$&	0.19334458&&	0.17815633&	0.17815294&&	0.14683800&	0.14683813&	0.14683852&&	0.09678618&	0.09678623&	0.09678639&	0.09678666\\
	&$10^{7}$&	0.19330816&&	0.17808454&	0.17803027&&	0.14675666&	0.14677158&	0.14681633&&	0.09672344&	0.09672894&	0.09674547&	0.09677302\\
$10^{21}$&	0&	0.03820810&&	0.02020779&	0.02020779&&&&&&&&	\\						
	&$10^{3}$&	0.03820777&&	0.02020659&	0.02020495&&&&&&&&	\\						
	&$10^{5}$&	0.03820756&&	0.02020497&	0.02020436&&&&&&&&	\\						
	&$10^{7}$&	0.03818578&&	0.02004326&	0.02014525&&&&&&&&	\\						
 \hline\hline
\end{tabular}
\end{scriptsize}
\end{center}
\end{sidewaystable}
For different sets of plasma density ($n_{e}$) and ion velocity ($v$), table 2-5 displays the energy eigenvalues of $n=1$ to $n=4$ states respectively. The energy eigenvalues of free ions are given in the first row of each table. From the numbers quoted in the tables 2-5, it is to be noted that as the plasma density ($n_{e}$) increases for a given ion velocity ($v$), the energy eigenvalues become more and more positive leading towards destabilization of the ion while as the ion velocity increases for a given plasma density ($n_{e}$), the energy eigenvalue becomes more and more positive but with a much slower rate compared to the preceding one. Thus it can be argued that the effect of static screening (depends only on $n_{e}$) of the plasma environment on the energy eigenvalue is much more pronounced compared to the wake field, where the later arises due to the velocity ($v$) of the ion and also depends on plasma electron density ($n_{e}$). In contrast, Hu \textit{et. al.} [32] showed that the energy of each state considered here becomes over-bound (\textit{i.e.} more negative than the energy of the free ion) when the ion velocity ($v$) reaches a sufficiently high value. For example, Hu \textit{et. al.} [32] reported that for $n_{e}=8.0\times 10^{23}$ m$^{-3}$ and ion velocity $v$ = 5000 m/s, the ground state ($1s_{0}$) energy of C$^{5+}$ becomes -25.60524 a.u. which is more negative than the ground state energy -18.0 a.u. for the free C$^{5+}$ ion. In this regard, Hu \textit{et. al.} [32] opined that such over-boundness occurred because of the choice of angular part of the wave function. But the angular part of the wavefunction cannot be responsible for such over-boundness as it violets the basic variational principle. The error lies in the deduction of the form of the near-field wake potential. Moreover, the -\textit{ve} sign for the near-field wake potential as obtained by Hu \textit{et. al.} [32] is not correct. No such over-boundness are being observed in the present calculations, \textit{e.g.}, we have obtained the ground state ($1s_{0}$) energy of -17.33679 a.u. for C$^{5+}$ ion, where $n_{e}=8.0\times 10^{23}$ m$^{-3}$ and $v$ = 5000 m/s.\\
It can also be noted from the tables 3-5 that the usual breaking of accidental degeneracy (\textit{i.e.} $l$ degeneracy corresponding to a given $n$) occurs \textit{w.r.t.} the plasma electron density ($n_{e}$). This is a well-known phenomenon in presence of Debye-Huckel potential and can be found in different studies [15]. The degeneracy of energy eigenvalues \textit{w.r.t.} the absolute value of the magnetic quantum number \textit{i.e.} $\vert m\vert$ is being removed for each ion velocity ($v$) because of the presence of $\cos\theta$ term in the near-field wake potential as seen in the tables 3-5. For example, table-3 shows that for ion velocity ($v$) $10^{3}$ c.m./sec. and plasma electron density ($n_{e}$) $10^{19}$/c.c., the energy eigenvalues of $2p_{0}$ and $2p_{1}$ states are $-3.54472655$ a.u. and $-3.54475747$ a.u. respectively. This is purely Stark-like splitting. In contrast, Hu \textit{et. al.} [32] reported the lifting of degeneracy of the energy levels \textit{w.r.t} magnetic quantum no. `$m$' \textit{i.e.} Zemman-like splitting. The error lies in the calculation of the matrix element of kinetic energy as given in the expression of $G(\beta,\gamma)$ of appendix-B of ref-[32]. In deriving $G(\beta,\gamma)$, Hu \textit{et. al.} [32] used the condition that the magnetic quantum number \textit{m} must be non-negative while in the numerical calculation of the energy eigenvalue they have used -\textit{ve} values of \textit{m}.\\
The variation of energies of ($2p_{0}$, $2p_{1}$),($3p_{0}$, $3p_{1}$), ($4p_{0}$, $4p_{1}$), ($3d_{0}$,$3d_{1}$,$3d_{2}$), ($4d_{0}$,$4d_{1}$,$4d_{2}$) and ($4f_{0}$,$4f_{1}$,$4f_{2}$,$4f_{3}$) states with ion velocity ($v$) for four different plasma densities ($n_{e}$) are depicted in figure 1-6 respectively. It is evident from figures 1-3 that corresponding to plasma density ($n_{e}$) $10^{19}$/c.c, $np_{1}$ states energetically lies below $np_{0}$ state for the entire range of ion velocity ($v$) and thus no crossing of energy levels are being observed. But for other three densities higher than the previous one, it is to be noted that, for low ion velocity $np_{0}$ state energetically lies below than that of $np_{1}$ state while after a critical ion velocity $np_{1}$ state becomes more negative than that of $np_{0}$ state. Hence, incidental degeneracy of $np_{0}$ and $np_{1}$ states occurs at the critical ion velocity. Such crossing of energy levels and subsequent appearance of incidental degeneracy occur for all other angular momentum states (\textit{i.e.} \textit{d}, \textit{f}) as given in figures 4-6. Such incidental degeneracy was reported earlier by Sen [42] in case of shell confined hydrogen atom. Thus the figures 1-6 give a good insight on the combined effect of static screening and near field wake potential on different angular momentum states of a slowly moving ion in quantum plasma.\\
\section{\label{sec:level2}Conclusion}
The electrostatic potential for a moving ion under quantum plasma are being derived where the thermal Mac number remains below unity. Subsequently, the effect of such potential on the change of the energy eigenvalues of different states of hydrogen-like carbon ion are being studied under the framework of Rayleigh-Ritz variational method. Level crossing phenomenon and incidental degeneracy are being observed for the first time in case of ion moving in the quantum plasma environment. The present form of the potential will help future workers to investigate the structural properties of different ions under quantum plasma environment.
\begin{center}
\textbf{Acknowledgment}
\end{center}
Authors express heartfelt thanks to Prof. H. Hu for helpful discussions. TKM gratefully acknowledges financial support under grant number 37(3)/14/27/2014-BRNS from the Department of Atomic Energy, BRNS, Government of India.

\end{document}